\documentclass[twocolumn, preprintnumbers, nofootinbib, superscriptaddress]{revtex4-2}

\usepackage{amsmath, amssymb}
\usepackage{graphicx}
\usepackage{hyperref}
\usepackage{xcolor}
\usepackage{booktabs}
\usepackage{multirow}
\usepackage{siunitx}
\usepackage{cancel}

\newcommand{\rev}[1]{\textcolor{black}{#1}}  
\newcommand{\GeV}{\,\text{GeV}}
\newcommand{\MeV}{\,\text{MeV}}
\newcommand{\fb}{\,\text{fb}}
\newcommand{\pb}{\,\text{pb}}
\newcommand{\MA}{M_A}
\newcommand{\gA}{g_A}
\newcommand{\GF}{G_F}
\newcommand{\Vud}{V_{ud}}
\newcommand{\slashed}[1]{\cancel{#1}}
\newcommand{\order}[1]{\mathcal{O}(#1)}

\newcommand{\CC}{\text{CC}}
\newcommand{\NC}{\text{NC}}

\newcommand{\LH}{\text{LH}}
\newcommand{\RH}{\text{RH}}
\newcommand{\Wminus}{W^-}

\begin{document}

\title{Constraining Neutrino--Nucleon Form Factors \\
with Charged-Current Scattering at the Electron-Ion Collider}

\author{Guang Yang}
\affiliation{Brookhaven National Laboratory, Upton, NY 11973, USA}

\author{Praveen Kumar}
\affiliation{Brookhaven National Laboratory, Upton, NY 11973, USA}
\affiliation{University of Alabama, Tuscaloosa, AL 35487, USA}

\date{\today}
\begin{abstract}
Next-generation neutrino oscillation experiments such as DUNE require
percent-level knowledge of neutrino--nucleon interaction cross sections.
The nucleon axial form factor $F_A(Q^2)$, parameterized by the axial mass
$\MA$, is the dominant source of uncertainty in the quasi-elastic channel,
and the parity-violating structure function $xF_3$ is poorly constrained
on free nucleons.
We propose using charged-current (CC) electron--proton scattering at the
Electron-Ion Collider (EIC) to address both problems simultaneously.
The measurement exploits three key features of the EIC:
(1)~helicity-selective electron bunches provide \emph{in situ}
electromagnetic background rejection;
(2)~a longitudinally polarized proton target enables extraction of
$F_A(Q^2)$ through the target-spin asymmetry $A_{UL}$; and
(3)~the $y$-distribution leverage in CC deep inelastic scattering
separates $F_2$ and $xF_3$ on a \emph{free proton}, without nuclear
corrections.
Using a Fisher-information analysis at $\sqrt{s} = 141\GeV$ with
$500\fb^{-1}$ of integrated luminosity, we project the Cram\'{e}r--Rao
statistical floor of $\delta\MA \approx 0.03\GeV$ (3\%).
Incorporating first-order realistic detector effects: ZDC acceptance,
$Q^2$ smearing (5\%), and background noise from helicity
subtraction, the projected sensitivity is severely background-limited
due to the small signal-to-background ratio ($S/B \approx 3 \times 10^{-4}$)
in the elastic channel.
Achieving competitive sensitivity ($\delta\MA \approx 0.14\GeV$)
would require $\sim\!10^{-7}$ background suppression, three orders
of magnitude beyond current projections.
The CC DIS $y$-distribution provides sub-percent extraction of
$xF_3^{\Wminus}$ over $0.05 < x < 0.5$, representing the most
robust electroweak measurement in the near term.
\end{abstract}

\maketitle

\section{Introduction}
\label{sec:intro}

The Deep Underground Neutrino Experiment (DUNE)~\cite{DUNE_TDR} aims to
measure the CP-violating phase $\delta_{CP}$ in the neutrino mixing matrix
with a precision of $\sim\!5^\circ$, determine the neutrino mass ordering
at $>5\sigma$, and search for proton decay and supernova neutrinos.
Achieving these physics goals requires knowledge of neutrino--nucleus
interaction cross sections at the few-percent level~\cite{Formaggio2012}.
At the relevant energies ($E_\nu \sim 1$--$5\GeV$), an important
interaction channel is charged-current quasi-elastic (CCQE) scattering:
\begin{equation}
\nu_\mu + n \to \mu^- + p\,.
\label{eq:ccqe}
\end{equation}

The CCQE cross section is governed by four nucleon form factors: the
vector Dirac ($F_1$) and Pauli ($F_2$) form factors, the axial form factor
$F_A$, and the pseudoscalar form factor $F_P$.
Through the conserved vector current (CVC) hypothesis, $F_1$ and $F_2$
are constrained to a few percent precision from decades of electron--nucleon
elastic scattering~\cite{Arrington2007}.
The pseudoscalar $F_P$ is related to $F_A$ by the partially conserved
axial current (PCAC) relation and contributes proportionally to $m_\ell^2/M^2$,
which is negligible for electrons ($\sim 10^{-7}$) and small for muons ($\sim 1\%$).
The least precisely constrained ingredient is the \emph{axial form factor}~$F_A(Q^2)$.

\subsection{The axial mass anomaly}
\label{sec:anomaly}

The standard parameterization of $F_A$ is the dipole form:
\begin{equation}
F_A(Q^2) = \frac{\gA}{(1 + Q^2/\MA^2)^2}\,,
\label{eq:FA_dipole}
\end{equation}
where $\gA = F_A(0) = 1.2723 \pm 0.0023$ is precisely known from neutron
beta decay~\cite{PDG2024}, and $\MA$ is the \emph{axial mass} that
controls the $Q^2$ dependence.
The axial charge radius is $\langle r_A^2\rangle = 12/\MA^2$.

\begin{figure}[t]
\centering
\includegraphics[width=\columnwidth]{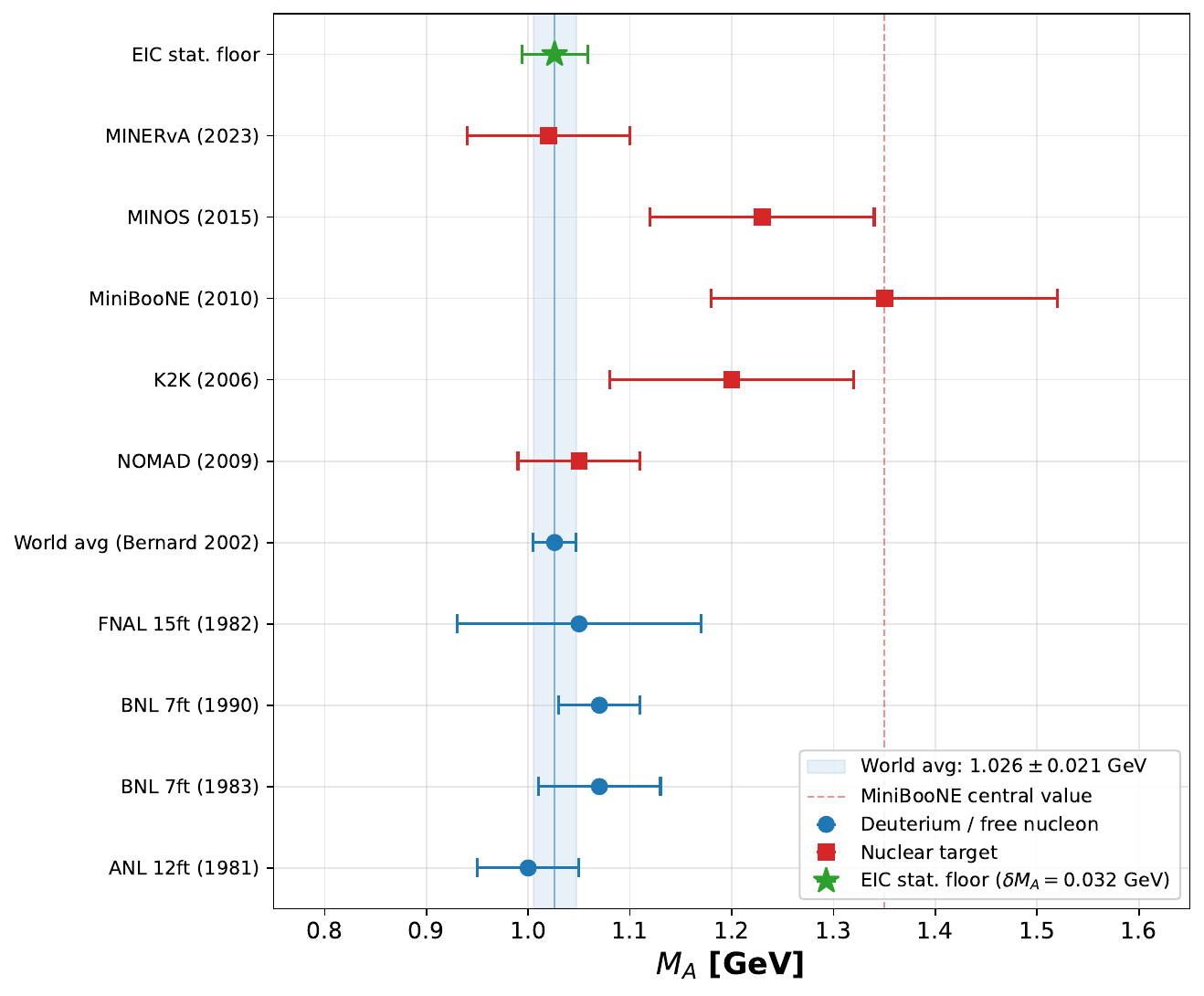}
\caption{Published measurements of the nucleon axial mass $\MA$
(data compiled from Refs.~\cite{Formaggio2012, BernardMA}).
Blue circles: measurements on deuterium (quasi-free nucleon).
Red squares: measurements on nuclear targets.
Green star: EIC statistical floor ($\delta\MA = 0.032\GeV$;
see Table~\ref{tab:fisher_results}).
The light blue band shows the world average
$\MA = 1.026 \pm 0.021\GeV$~\cite{BernardMA}; the dashed red line marks
the MiniBooNE central value~\cite{MiniBooNE_MA}.}
\label{fig:MA_anomaly}
\end{figure}

Measurements of $\MA$ have yielded conflicting results
(Fig.~\ref{fig:MA_anomaly}):
\begin{itemize}
\item \emph{Deuterium bubble chambers} (1970s--80s):
$\MA = 1.026 \pm 0.021\GeV$~\cite{BernardMA}, from quasi-free nucleon
targets with minimal nuclear corrections.
\item \emph{MiniBooNE} (2010): $\MA = 1.35 \pm 0.17\GeV$ on
carbon~\cite{MiniBooNE_MA}, a $\sim\!2\sigma$ excess over the world
average.
\item \emph{MINERvA} (2023): The first measurement on
hydrogen~\cite{MINERvA_Nature}, but with limited statistics and
reliance on flux modeling.
\end{itemize}

The discrepancy between the bubble chamber and MiniBooNE values,
often called the ``$\MA$ anomaly,'' has been attributed to nuclear effects in carbon
(multi-nucleon correlations, meson exchange currents, random phase
approximation corrections)~\cite{Martini2009, Nieves2011},
but the situation remains unresolved.
A high-precision measurement of $\MA$ on a \emph{free proton} would
definitively establish the nucleon-level form factor and disentangle
it from nuclear effects.

\subsection{The EIC opportunity}
\label{sec:eic_opportunity}

The Electron-Ion Collider~\cite{EIC_YR}, under construction at Brookhaven
National Laboratory, will collide longitudinally polarized electrons
($E_e = 5$--$18\GeV$) with longitudinally polarized protons
($E_p = 41$--$275\GeV$).
At the maximum energy configuration ($18\GeV \times 275\GeV$), the
center-of-mass energy is $\sqrt{s} = 141\GeV$, with a design luminosity
of $\mathcal{L} = 10^{34}\,\text{cm}^{-2}\text{s}^{-1}$.

As pointed out by Klest~\cite{Klest2025}, the EIC can access
charged-current elastic scattering:
\begin{equation}
e^- + p \to \nu_e + n\,,
\label{eq:cce}
\end{equation}
which is related to the neutrino CCQE process~\eqref{eq:ccqe} by
crossing symmetry.
The key advantage is that the EIC provides a \emph{free proton target}
with \emph{controlled beam properties}: known energy, polarization,
and luminosity, none of which are available in neutrino experiments.

In this paper, we propose a three-phase measurement program:
\begin{enumerate}
\item \textbf{Phase~1: Helicity filtering} (Sec.~\ref{sec:phase1}).
Using right-handed (RH) electron bunches as an \emph{in situ}
electromagnetic background template.
\item \textbf{Phase~2: Axial form factor extraction}
(Sec.~\ref{sec:phase2}).
Measuring the $Q^2$-binned cross section and target-spin asymmetry
$A_{UL}$ to determine $\MA$.
\item \textbf{Phase~3: Structure function extraction}
(Sec.~\ref{sec:phase3}).
Using the CC DIS $y$-distribution to separate $F_2^{\Wminus}$ and
$xF_3^{\Wminus}$ on a free proton.
\end{enumerate}

We quantify projected sensitivities using the Fisher information
formalism~\cite{Cowan1998}, which gives the Cram\'{e}r--Rao lower
bound on parameter uncertainties.
For the ideal projection, detector resolution, reconstruction efficiencies,
and systematic uncertainties are not included; these results represent
the statistical floor of the measurement.
A realistic projection incorporating first-order detector effects is
also presented.
The purpose is to establish whether the EIC has sufficient statistical
reach to make a competitive measurement, not to perform a full
experimental analysis.

\subsection{Outline}

Section~\ref{sec:theory} reviews the theoretical framework.
Section~\ref{sec:setup} describes the EIC experimental setup.
Section~\ref{sec:method} presents the sensitivity projection
methodology, including all assumptions.
Sections~\ref{sec:phase1}--\ref{sec:phase3} present the three analysis
phases and their results.
Section~\ref{sec:discussion} discusses systematics, nuclear targets,
and comparisons with neutrino experiments.
Section~\ref{sec:conclusions} summarizes the conclusions.

\section{Theoretical Framework}
\label{sec:theory}

\subsection{Charged-current elastic scattering}
\label{sec:elastic_theory}

The matrix element for the CC elastic process
$e^-(k) + p(P) \to \nu_e(k') + n(P')$ is:
\begin{equation}
\mathcal{M} = \frac{\GF \Vud}{\sqrt{2}}\,
\bar{u}(k')\, \gamma^\mu (1 - \gamma_5)\, u(k)\;
\bar{u}(P')\, \Gamma_\mu\, u(P)\,,
\label{eq:matrix_element}
\end{equation}
where the hadronic vertex function is:
\begin{equation}
\Gamma_\mu = \gamma_\mu F_1(Q^2)
+ \frac{i\sigma_{\mu\nu}q^\nu}{2M} F_2(Q^2)
+ \gamma_\mu\gamma_5 F_A(Q^2)
+ \frac{q_\mu}{M}\gamma_5 F_P(Q^2)\,.
\label{eq:vertex}
\end{equation}
Here $q = k - k' = P' - P$ is the four-momentum transfer,
$Q^2 = -q^2 > 0$, and $M = (M_p + M_n)/2 = 0.939\GeV$ is the average
nucleon mass.

\subsubsection{Form factors}
\label{sec:form_factors}

The \emph{vector form factors} $F_1$ and $F_2$ are determined by CVC:
\begin{equation}
F_{1,2}^V(Q^2) = F_{1,2}^p(Q^2) - F_{1,2}^n(Q^2)\,.
\end{equation}
These are expressed in terms of the Sachs form factors via
\begin{align}
F_1 &= \frac{G_E^V + \tau G_M^V}{1 + \tau}\,, &
F_2 &= \frac{G_M^V - G_E^V}{1 + \tau}\,,
\label{eq:sachs_to_dirac}
\end{align}
where $\tau = Q^2/(4M^2)$.
We use the standard dipole parameterization for the proton,
\begin{equation}
G_E^p = G_D \equiv \frac{1}{(1 + Q^2/M_V^2)^2}\,, \quad
G_M^p = \mu_p\, G_D\,,
\end{equation}
with $M_V^2 = 0.71\GeV^2$ and $\mu_p = 2.793$~\cite{PDG2024},
the Galster parameterization for the neutron,
\begin{equation}
G_E^n = -\frac{\mu_n\,\tau}{1 + 5.6\,\tau}\,G_D\,,
\end{equation}
and dipole scaling $G_M^n = \mu_n\,G_D$ with
$\mu_n = -1.913$~\cite{PDG2024}.
These are known to 1--2\% precision~\cite{Arrington2007} and are treated as fixed inputs
throughout this analysis.

The \emph{axial form factor} $F_A(Q^2)$ is given by
Eq.~\eqref{eq:FA_dipole}, with the two free parameters $\gA$ and $\MA$.
The \emph{pseudoscalar form factor} follows from PCAC
(pion-pole dominance):
\begin{equation}
F_P(Q^2) = \frac{2M^2\, F_A(Q^2)}{Q^2 + m_\pi^2}\,.
\label{eq:PCAC}
\end{equation}
For massless leptons, $F_P$ contributes proportional to $m_\ell^2/M^2$
and is negligible.

\subsubsection{Llewellyn Smith formula}
\label{sec:llewellyn_smith}

The unpolarized differential cross section is~\cite{LlewellynSmith1972}:
\begin{equation}
\frac{d\sigma}{dQ^2} = \frac{M^2\,\GF^2\,|\Vud|^2}{8\pi E_\nu^2}
\left[A - B\,\frac{s-u}{M^2} + C\,\frac{(s-u)^2}{M^4}\right],
\label{eq:llewellyn_smith}
\end{equation}
where $E_\nu = (s - M^2)/(2M)$ is the equivalent neutrino energy in
the target rest frame and $s - u = 4ME_\nu - Q^2$.
The minus sign before $B$ corresponds to $W^-$ exchange (antineutrino
sign convention).
The coefficients in the massless-lepton limit are:
\begin{align}
A &= \frac{Q^2}{M^2}\!\left[(1\!+\!\tau)F_A^2 - (1\!-\!\tau)F_1^2
+ \tau(1\!-\!\tau)F_2^2 + 4\tau F_1 F_2\right]\!,
\label{eq:A_coeff} \\
B &= \frac{Q^2}{M^2}\,F_A(F_1 + F_2)\,,
\label{eq:B_coeff} \\
C &= \frac{1}{4}\left(F_A^2 + F_1^2 + \tau F_2^2\right).
\label{eq:C_coeff}
\end{align}

We emphasize that the \emph{only unknowns} in
Eq.~\eqref{eq:llewellyn_smith} are $\gA$ and $\MA$, since $F_1$ and
$F_2$ are fixed by CVC and $F_P$ is negligible.
The total cross section at EIC energies is
$\sigma_\text{el} \approx 9.3\fb$ for $\MA = 1.026\GeV$.

\subsubsection{Target-spin asymmetry $A_{UL}$}
\label{sec:AUL_theory}

With a longitudinally polarized proton target, the target-spin
asymmetry is:
\begin{equation}
A_{UL}(Q^2) = \frac{d\sigma(\uparrow) - d\sigma(\downarrow)}
{d\sigma(\uparrow) + d\sigma(\downarrow)}\,,
\label{eq:AUL_def}
\end{equation}
where $\uparrow$/$\downarrow$ denotes proton spin aligned/anti-aligned
with the beam.
$A_{UL}$ arises from V--A interference and depends on the relative
magnitudes of $F_A$ and the vector form factors.
At $Q^2 \to 0$, the axial coupling dominates and $|A_{UL}| \to 1$.
At larger $Q^2$, the vector contribution grows and the asymmetry
decreases; the rate of decrease depends on $\MA$.
We compute $A_{UL}$ numerically using the Dirac trace technique
(Appendix~\ref{app:trace}).

\subsubsection{Crossing symmetry}
\label{sec:crossing}

The EIC process~\eqref{eq:cce} is related to the neutrino CCQE
process~\eqref{eq:ccqe} by crossing symmetry. Both probe the
\emph{same} hadronic vertex~\eqref{eq:vertex} with the \emph{same}
form factors. The cross sections differ only in kinematics and in the
sign of the $B$ coefficient. A measurement of $\MA$ at the EIC
therefore directly constrains the neutrino CCQE cross section.

\subsection{CC deep inelastic scattering}
\label{sec:dis_theory}

At $x < 1$, the electron scatters off individual quarks inside the
proton. The CC DIS cross section is:
\begin{equation}
\frac{d^2\sigma^{\CC}}{dx\,dy} = \frac{\GF^2\,s}{4\pi}
\left(\frac{M_W^2}{Q^2 + M_W^2}\right)^{\!2}
\left[Y_+ F_2 - Y_- xF_3 - y^2 F_L\right],
\label{eq:cc_dis}
\end{equation}
where $Q^2 = xys$ and:
\begin{equation}
Y_+ = 1 + (1-y)^2\,, \qquad Y_- = 1 - (1-y)^2 = 2y - y^2\,.
\label{eq:Y_coeffs}
\end{equation}

For $W^-$ exchange on the proton ($e^- p \to \nu_e X$), the
leading-order structure functions are:
\begin{align}
F_2^{W^-}(x) &= 2x\left[u(x) + c(x) + \bar{d}(x) + \bar{s}(x)\right],
\label{eq:F2_Wminus} \\
xF_3^{W^-}(x) &= 2x\left[u(x) + c(x) - \bar{d}(x) - \bar{s}(x)\right].
\label{eq:xF3_Wminus}
\end{align}
Thus $xF_3$ is sensitive to the difference between quark and antiquark
distributions, dominated by valence quarks at moderate $x$.
These are the same structure functions probed by
$\bar{\nu}_\mu p \to \mu^+ X$ (both involve $W^-$ exchange), but the
EIC measurement is on a \emph{free proton} with known beam energy.

The key observation is that $F_2$ and $xF_3$ multiply different
functions of $y$: at $y \to 0$, $Y_+ \to 2$ but $Y_- \to 0$, so only
$F_2$ contributes; at $y \to 1$, both contribute equally.
By fitting the $y$-distribution at fixed $x$, one extracts both
$F_2(x)$ and $xF_3(x)$ simultaneously.
We set $F_L = 0$ (Callan--Gross relation), valid at leading order;
NLO corrections give $F_L \sim \order{\alpha_s/(4\pi)} \times F_2$,
suppressed by the $y^2$ prefactor.

\section{EIC Experimental Setup}
\label{sec:setup}

\subsection{Machine parameters}

We assume the EIC at its maximum energy: $E_e = 18\GeV$ colliding with
$E_p = 275\GeV$, giving $\sqrt{s} = \sqrt{4 E_e E_p} = 140.7\GeV$
and $s = 19{,}800\GeV^2$.
The equivalent fixed-target energy is
$E_e^* = s/(2M_p) \approx 10{,}554\GeV$.
The parameters relevant to this analysis are summarized in
Table~\ref{tab:eic_params}.

\begin{table}[t]
\centering
\caption{EIC parameters used in this analysis. All values from the
EIC Yellow Report~\cite{EIC_YR} unless otherwise noted.}
\label{tab:eic_params}
\begin{tabular}{lll}
\toprule
Parameter & Value & Note \\
\midrule
$E_e$ & 18 GeV & Maximum \\
$E_p$ & 275 GeV & Maximum \\
$\sqrt{s}$ & 140.7 GeV & \\
$\mathcal{L}$ & $10^{34}$ cm$^{-2}$s$^{-1}$ & Design \\
$\int\!\mathcal{L}\,dt$ & 500 fb$^{-1}$ & 5 years \\
Electron pol.\ $P_e$ & 80\% & Longitudinal \\
Proton pol.\ $P_p$ & 70\% & Longitudinal \\
Polarimetry unc.\ & 1\% (relative) & \\
\bottomrule
\end{tabular}
\end{table}

\begin{figure}[t]
\centering
\includegraphics[width=\columnwidth]{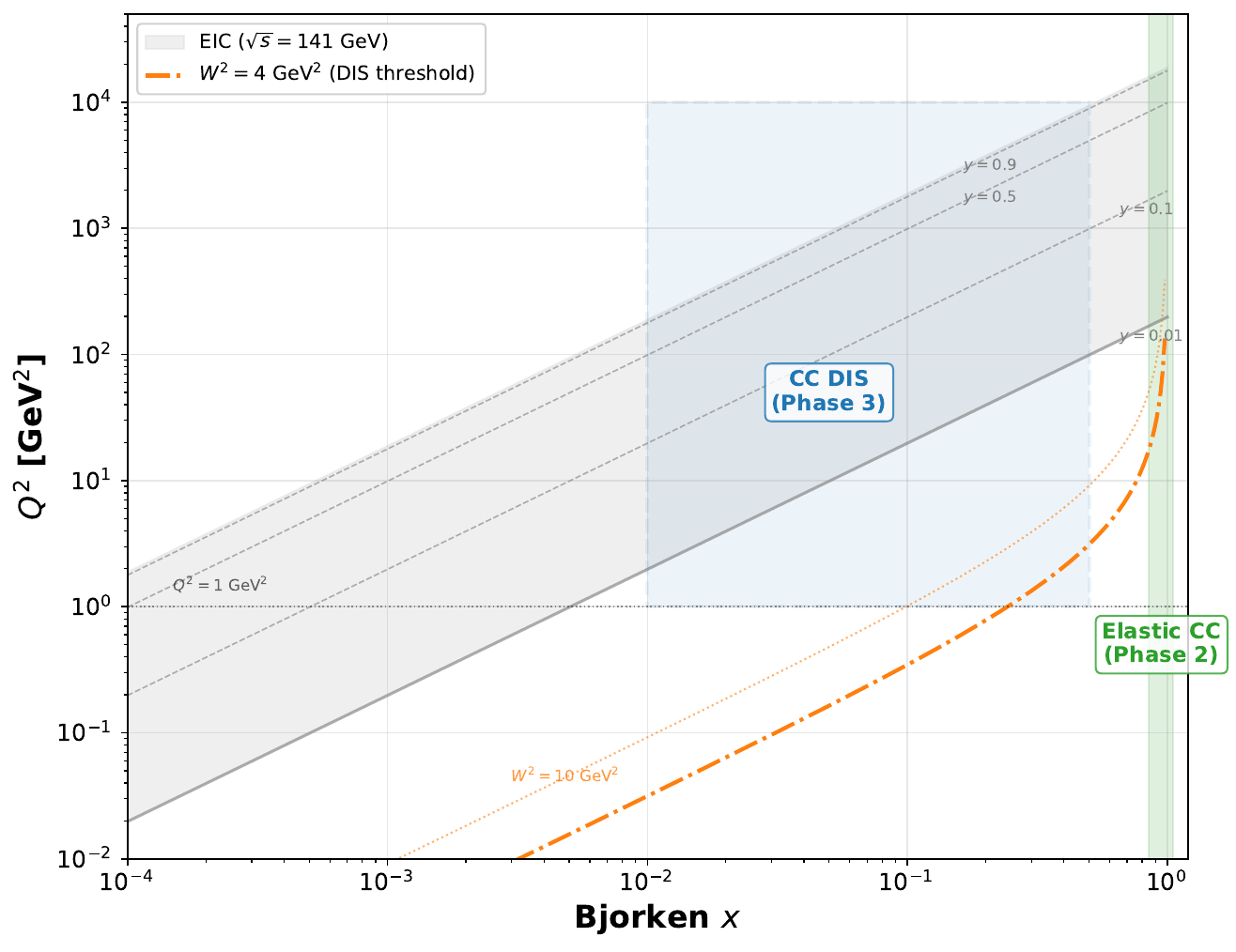}
\caption{Kinematic coverage of the EIC at $\sqrt{s} = 141\GeV$ in the
$(x, Q^2)$ plane. Gray shading: accessible phase space for
$0.01 < y < 0.95$. Green band: elastic CC region ($x \approx 1$).
Blue dashed rectangle: CC DIS region used for $xF_3$ extraction.
Orange dot-dashed line: $W^2 = 4\GeV^2$ (DIS threshold).
Diagonal gray lines: constant $y$.}
\label{fig:kinematics}
\end{figure}

The kinematic coverage is shown in Fig.~\ref{fig:kinematics}.
The elastic CC channel corresponds to $x = 1$, $Q^2 \sim 0.01$--$3\GeV^2$
(Phase~2), while CC DIS at $Q^2 > 1\GeV^2$, $0.01 < x < 0.5$ provides
the $y$-leverage for structure function extraction (Phase~3).

\subsection{Signal topology and backgrounds}
\label{sec:signal_bg}

The elastic CC event $e^- p \to \nu_e n$ produces a forward neutron
(detected in the Zero Degree Calorimeter, ZDC) and a neutrino (missing
energy), with no scattered electron in the central detector.
The signature is a leading neutron with $x_L = E_n/E_p > 0.88$ and no
other activity, as studied by Klest~\cite{Klest2025}.

The dominant background is \emph{leading neutron photoproduction}:
$\gamma^* p \to n \pi^+ X$, with a cross section
$\sigma_\text{bg} \sim 0.3\,\mu\text{b}$~\cite{Klest2025}, roughly
$3 \times 10^4$ times larger than the CC elastic signal.
Background rejection comes from:
(i)~a forward charged-particle veto (three silicon layers at 99.8\%
efficiency each)~\cite{Klest2025};
(ii)~the $x_L > 0.88$ cut; and
(iii)~helicity filtering (Sec.~\ref{sec:phase1}).
After the veto and kinematic cuts, we estimate a combined
background suppression factor of $\sim\!10^{-4}$, yielding a
residual background cross section
$\sigma_\text{bg}^\text{after} \approx 30{,}000\fb$
($= 0.3\,\mu\text{b} \times 10^{-4}$, using $1\,\mu\text{b} = 10^9\fb$).

\section{Sensitivity Projection Methodology}
\label{sec:method}

All sensitivity projections in this paper use the Fisher information
matrix formalism~\cite{Cowan1998}. This section describes the method,
inputs, and assumptions in detail.

\subsection{Fisher information matrix}
\label{sec:fisher_formalism}

For a counting experiment with Poisson-distributed event counts
$N_k$ in bins $k$ that depend on parameters $\theta_i$, the Fisher
information matrix is:
\begin{equation}
\mathcal{F}_{ij} = \sum_k \frac{1}{N_k(\boldsymbol{\theta})}
\frac{\partial N_k}{\partial\theta_i}\,
\frac{\partial N_k}{\partial\theta_j}\,.
\label{eq:fisher_general}
\end{equation}
The covariance matrix of the maximum-likelihood estimator is bounded
from below by $C \geq \mathcal{F}^{-1}$ (the Cram\'{e}r--Rao
inequality), so $\delta\theta_i \geq \sqrt{(\mathcal{F}^{-1})_{ii}}$.
Our projections report this lower bound.
\rev{The Fisher approach assumes the Gaussian limit of Poisson
statistics (valid for $N_k \gtrsim 10$) and linear parameter
dependence near the minimum. With $\sim\!1{,}440$ total events
across 11 bins, most bins have $N_k \sim 50$--$300$, satisfying
the Gaussian assumption. For a complete experimental analysis, a
binned maximum-likelihood fit with Monte Carlo pseudo-experiments
would be required to validate the Fisher projections; the purpose
here is to establish the statistical feasibility of the measurement.}

When an \emph{asymmetry} $A_k$ is also measured in each bin with
Gaussian uncertainty $\delta A_k$, an additional Fisher contribution
is:
\begin{equation}
\mathcal{F}_{ij}^{(A)} = \sum_k \frac{1}{(\delta A_k)^2}\,
\frac{\partial A_k}{\partial\theta_i}\,
\frac{\partial A_k}{\partial\theta_j}\,.
\label{eq:fisher_asym}
\end{equation}
The total Fisher matrix is
$\mathcal{F} = \mathcal{F}^{(\sigma)} + \mathcal{F}^{(A)}$.

\subsection{Inputs and assumptions}
\label{sec:assumptions}

The following assumptions apply throughout:

\paragraph{Event counting.}
The expected number of elastic CC events in $Q^2$ bin $k$ is:
\begin{equation}
N_k = \tfrac{1}{2}\,\mathcal{L}_\text{int}\,
\frac{d\sigma}{dQ^2}\bigg|_{Q^2_k} \Delta Q^2_k\,,
\label{eq:event_counting}
\end{equation}
where the factor $1/2$ accounts for the fact that only left-handed
electrons produce CC interactions (Sec.~\ref{sec:phase1}), and we
assume 50\% of the delivered luminosity is in each helicity state.
The cross section $d\sigma/dQ^2$ is evaluated at the bin center
using the Llewellyn Smith formula~\eqref{eq:llewellyn_smith} with
fixed inputs from Table~\ref{tab:inputs}.

\begin{table}[t]
\centering
\caption{Fixed physics inputs used in the sensitivity projections.}
\label{tab:inputs}
\begin{tabular}{lll}
\toprule
Quantity & Value & Source \\
\midrule
$\GF$ & $1.1664 \times 10^{-5}\GeV^{-2}$ & PDG~\cite{PDG2024} \\
$|\Vud|$ & 0.97373 & PDG~\cite{PDG2024} \\
$M$ (nucleon avg) & $0.939\GeV$ & PDG~\cite{PDG2024} \\
$M_W$ & $80.377\GeV$ & PDG~\cite{PDG2024} \\
$\gA$ (fiducial) & 1.2723 & PDG~\cite{PDG2024} \\
$\MA$ (fiducial) & $1.026\GeV$ & Ref.~\cite{BernardMA} \\
$M_V^2$ & $0.71\GeV^2$ & Electron scattering \\
$\mu_p$ & 2.793 & PDG~\cite{PDG2024} \\
$\mu_n$ & $-1.913$ & PDG~\cite{PDG2024} \\
$s$ & $19{,}800\GeV^2$ & EIC $18\times275$ \\
$\mathcal{L}_\text{int}$ & $500\fb^{-1}$ & 5 years \\
$P_p$ & 0.70 & EIC YR~\cite{EIC_YR} \\
\bottomrule
\end{tabular}
\end{table}

\paragraph{Asymmetry uncertainty.}
The statistical uncertainty on the target-spin asymmetry in each
$Q^2$ bin is:
\begin{equation}
\delta A_{UL,k} = \frac{1}{P_p \sqrt{N_k}}\,,
\label{eq:delta_AUL}
\end{equation}
where $P_p = 0.70$ is the proton beam polarization and $N_k$ is the
number of CC events in the bin.
This follows from $\text{Var}(A_\text{meas}) \approx 1/N$ for the
raw counting asymmetry, with the physical asymmetry diluted by the
polarization: $A_\text{meas} = P_p \cdot A_\text{true}$.

\paragraph{Numerical derivatives.}
Partial derivatives $\partial N_k/\partial\theta_i$ and
$\partial A_k/\partial\theta_i$ are computed by central finite
differences with a 1\% step size in each parameter (e.g.,
$\Delta\gA = 0.01 \times \gA$). We have verified that the results
are insensitive to the step size over the range 0.1--5\%.

\paragraph{Binning.}
For elastic CC: 11 logarithmically spaced $Q^2$ bins from 0.01 to
$3.0\GeV^2$.
For CC DIS: 19 uniform $y$ bins from 0.05 to 0.95, with $x$-bin
width $\Delta x = 0.05$.

\subsection{Realistic detector model}
\label{sec:detector_model}

In addition to the ideal (Cram\'{e}r--Rao) projections, we present
a first-order realistic estimate that incorporates detector effects.
The model parameters (Table~\ref{tab:detector}) are based on the
EIC Yellow Report~\cite{EIC_YR} and Klest~\cite{Klest2025}.

\begin{table}[t]
\centering
\caption{Detector model parameters for the realistic projection.}
\label{tab:detector}
\begin{tabular}{lll}
\toprule
Parameter & Value & Source \\
\midrule
\multicolumn{3}{l}{\emph{ZDC acceptance for elastic neutrons}} \\
Core acceptance ($\theta_n < 3.5$~mrad) & 85\% & YR~\cite{EIC_YR} \\
Edge falloff & 3.5--5.5~mrad & YR~\cite{EIC_YR} \\
\midrule
\multicolumn{3}{l}{\emph{Efficiency factors}} \\
ZDC trigger & 95\% & \\
$x_L > 0.88$ cut (signal) & 99\% & \\
Analysis quality cuts & 90\% & \\
\midrule
\multicolumn{3}{l}{\emph{Kinematic resolution}} \\
$\sigma(Q^2)/Q^2$ & 5\% & Ref.~\cite{Klest2025} \\
\midrule
\multicolumn{3}{l}{\emph{Systematic uncertainties}} \\
Luminosity & 2\% (relative) & YR~\cite{EIC_YR} \\
Polarimetry ($P_p$) & 1\% (relative) & YR~\cite{EIC_YR} \\
ZDC energy scale & 1\% (relative) & \\
LH/RH bg subtraction & 1\% of $N_\text{bg}$ & \\
\bottomrule
\end{tabular}
\end{table}

\paragraph{ZDC acceptance.}
The elastic neutron angle in the lab frame is
$\theta_n \approx \sqrt{Q^2}/E_p$. For $Q^2 = 1\GeV^2$,
$\theta_n \approx 3.6$~mrad; for $Q^2 = 2\GeV^2$,
$\theta_n \approx 5.1$~mrad. The ZDC has full efficiency
($\varepsilon_\text{ZDC} \approx 85\%$, accounting for shower
containment and dead material) for $\theta_n < 3.5$~mrad, linearly
falling to zero at $\theta_n = 5.5$~mrad. Combined with trigger
(95\%), $x_L$ cut (99\%), and analysis (90\%) efficiencies, the
total detection efficiency is $\varepsilon \approx 72\%$ for
$Q^2 \lesssim 0.9\GeV^2$, dropping to zero for $Q^2 \gtrsim 2.3\GeV^2$.
The effective number of detected CC events is $\sim\!1{,}440$
(compared to $\sim\!2{,}200$ in the ideal case).

\paragraph{$Q^2$ smearing.}
The $Q^2$ resolution of $\sim\!5\%$~\cite{Klest2025} causes
bin migration in the $Q^2$-binned cross section. We model this with a
Gaussian response matrix $R_{kj}$ where $R_{kj}$ is the probability
that a true-$Q^2$ event in bin~$j$ is reconstructed in bin~$k$.
The smeared expected events are $N_k^\text{reco} = \sum_j R_{kj}\,
N_j^\text{true}$.

\paragraph{Background in the Fisher matrix.}
After helicity subtraction~\eqref{eq:helicity_sub}, the expected
CC signal in each bin is unbiased, but the \emph{variance} is
$\sigma^2_k = S_k + 2B_k$ where $S_k$ is the signal and $B_k$ is
the background per helicity.
The Fisher information is therefore
$\mathcal{F}_{ij}^{(\sigma)} = \sum_k
(\partial S_k/\partial\theta_i)(\partial S_k/\partial\theta_j)
/ (S_k + 2B_k)$,
which is degraded by the factor $S_k/(S_k + 2B_k) \approx 0.07$--$0.21$
relative to the ideal case (varies by $Q^2$ bin).
Similarly, the asymmetry uncertainty including background dilution is
$\delta A_k = \sqrt{(S_k + 2B_k)} / (P_p\, S_k)$, degraded by
$\sqrt{1 + 2B_k/S_k}$ compared to the signal-only case.

\paragraph{Systematic uncertainties.}
For each source listed in Table~\ref{tab:detector}, we propagate
the assumed uncertainty to $\MA$ by computing the bias from a
$1\sigma$ shift:
\begin{itemize}
\item \emph{Luminosity} (2\%): scales the overall event count,
affecting the cross section normalization. Since $\MA$ is extracted
from the $Q^2$ shape, the impact is small: $\delta\MA = 0.004\GeV$.
\item \emph{ZDC energy scale} (1\%): shifts
$Q^2_\text{reco} \to Q^2_\text{true}(1 + \delta_E)$, distorting the
$Q^2$ shape. This is a significant systematic: $\delta\MA = 0.012\GeV$.
\item \emph{Polarimetry} (1\%): affects the extracted $A_{UL}$
asymmetry. Since the cross section shape dominates over $A_{UL}$ in
the realistic projection (due to background dilution), the impact is
small: $\delta\MA = 0.001\GeV$.
\item \emph{Background subtraction} (1\% LH/RH asymmetry): imperfect
helicity subtraction from LH/RH luminosity imbalance injects a
residual with the exponential background $Q^2$ shape. This is the
largest systematic: $\delta\MA = 0.015\GeV$.
\end{itemize}
The total systematic $\delta\MA^\text{syst} = 0.019\GeV$ (added in
quadrature) is subdominant to the statistical+detector uncertainty.
The systematic budget is summarized in Table~\ref{tab:syst_budget}
in Sec.~\ref{sec:systematics}.

\paragraph{What is \emph{not} included.}
The following effects are still omitted in both projections:
\begin{enumerate}
\item \emph{QCD corrections.} The elastic cross section is computed
at tree level. The DIS cross section uses leading-order structure
functions with $F_L = 0$ (Callan--Gross) and toy PDFs at a fixed
scale (Appendix~\ref{app:pdfs}). NLO QCD and DGLAP evolution
are omitted.
\item \emph{Radiative corrections.} QED corrections to the CC vertex
are $\order{\alpha/\pi} \sim 0.2\%$ and negligible.
\item \emph{Full detector simulation.} The response matrix and
efficiency models are parameterized, not from Geant4.
\item \emph{Unfolding.} We assume perfect bin-center corrections;
a real analysis would require iterative unfolding.
\end{enumerate}

\section{Phase 1: Helicity Filtering}
\label{sec:phase1}

\subsection{Principle}

The $W$ boson couples exclusively to left-handed fermions.
CC interactions can therefore only occur with LH electrons, while
electromagnetic processes occur with both helicities.
The EIC stores alternating bunches of longitudinally polarized electrons.
For an electron beam polarization $P_e = 0.80$:
\begin{itemize}
\item ``LH bunches'' contain $(1+P_e)/2 = 90\%$ LH and 10\% RH electrons;
\item ``RH bunches'' contain $(1-P_e)/2 = 10\%$ LH and 90\% RH electrons.
\end{itemize}
Recording data separately:
\begin{align}
N_\LH &= \tfrac{1+P_e}{2}\,N_\CC + N_\text{EM}\,, \\
N_\RH &= \tfrac{1-P_e}{2}\,N_\CC + N_\text{EM}\,,
\end{align}
the CC signal is extracted by polarization-corrected subtraction:
\begin{equation}
N_\CC = \frac{N_\LH - N_\RH}{P_e}\,,
\label{eq:helicity_sub}
\end{equation}
with statistical uncertainty
$\delta N_\CC = \sqrt{N_\LH + N_\RH}/P_e$.
The factor $1/P_e$ arises because the subtraction $N_\LH - N_\RH$
yields $(P_e) \cdot N_\CC$, not $N_\CC$ directly.
For $P_e = 0.80$, this correction factor is $1/0.80 = 1.25$, meaning:
(i)~the raw subtraction underestimates $N_\CC$ by 20\% if uncorrected,
and (ii)~the statistical uncertainty is amplified by the same factor,
$\delta N_\CC \propto 1/P_e$.
Higher polarization reduces this amplification and improves sensitivity.

\subsection{Background model and pseudo-data generation}
\label{sec:bg_model}

To illustrate the helicity filtering, we generate pseudo-data with
the following model:
\begin{itemize}
\item \emph{Signal:} CC elastic events from the Llewellyn Smith
formula~\eqref{eq:llewellyn_smith} with $\MA = 1.026\GeV$, counted
per Eq.~\eqref{eq:event_counting} with
$\mathcal{L}_\LH = 250\fb^{-1}$.
\item \emph{Background:} The residual photoproduction background
(after forward veto and $x_L$ cut) with total cross section
$\sigma_\text{bg}^\text{after} \approx 30{,}000\fb$
($= 0.3\,\mu\text{b} \times 10^{-4}$, using $1\,\mu\text{b} = 10^9\fb$).
The $Q^2$ shape is modeled as an exponential:
\begin{equation}
\frac{d\sigma_\text{bg}}{dQ^2} =
\frac{\sigma_\text{bg}^\text{after}}{\Lambda^2}\,
e^{-Q^2/\Lambda^2}\,, \quad \Lambda^2 = 0.3\GeV^2\,,
\end{equation}
reflecting the quasi-real photon origin ($Q^2 \to 0$).
This shape is an ansatz; the helicity subtraction removes it
regardless of the true shape.
\item \emph{Poisson sampling:} The expected event counts
$\langle N_k \rangle$ in each $Q^2$ bin are computed from the
cross section $\times$ luminosity. The \emph{observed} counts are
then drawn from Poisson distributions:
$N_k \sim \text{Poisson}(\langle N_k \rangle)$.
This standard Monte Carlo procedure produces statistical fluctuations
around the true expectation, representing what an actual experiment
would observe. In Fig.~\ref{fig:helicity}, the deviations of the
black data points from the green histogram reflect these Poisson
fluctuations.
\end{itemize}

\subsection{Results}

\begin{figure}[t]
\centering
\includegraphics[width=\columnwidth]{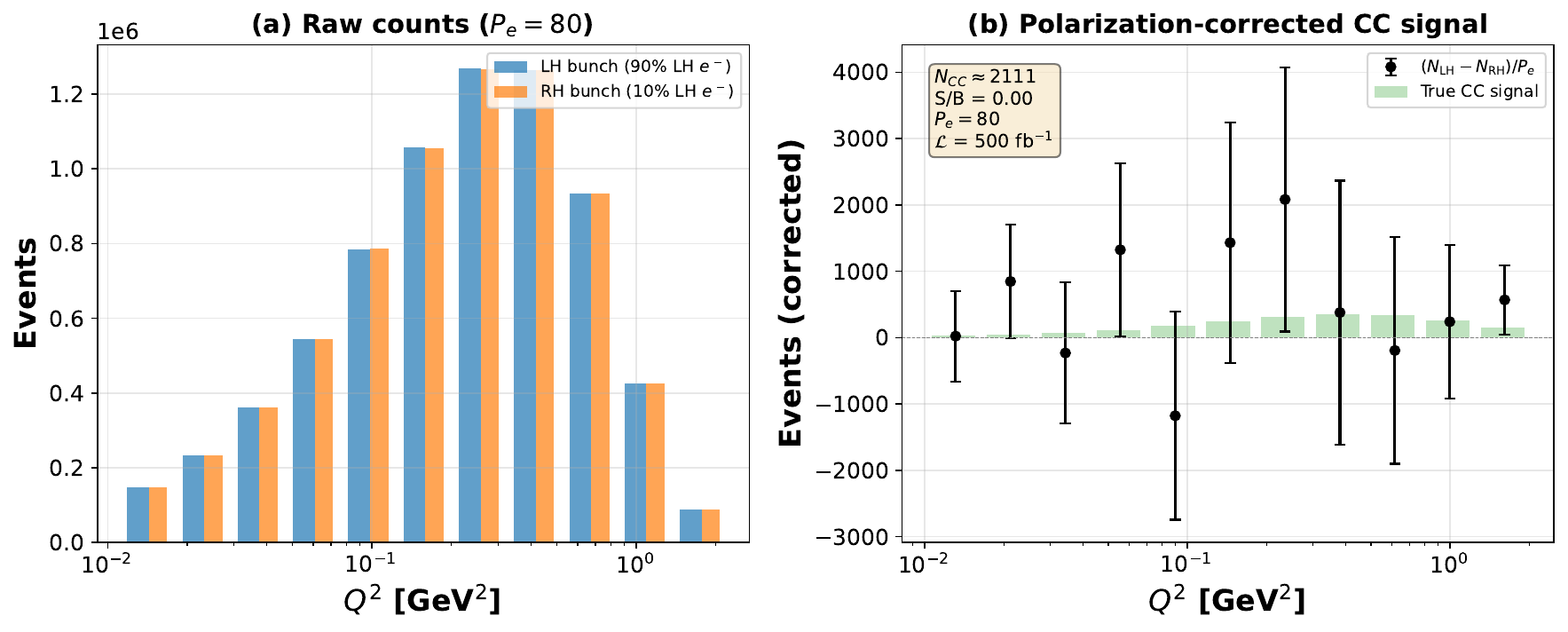}
\caption{Helicity filtering for CC elastic signal extraction
($500\fb^{-1}$ total, $250\fb^{-1}$ per helicity, $P_e = 0.80$).
\emph{Left:} Raw event counts by electron helicity. With 80\%
polarization, LH bunches contain 90\% LH electrons and RH bunches
contain 10\% LH electrons, so both produce CC events.
\emph{Right:} Polarization-corrected CC signal
$(N_\LH - N_\RH)/P_e$ (black points with error bars) compared to
the true CC signal (green bars). The error bars include the
$1/P_e$ amplification factor per Eq.~\eqref{eq:helicity_sub}.
Statistical fluctuations arise from Poisson sampling
(Sec.~\ref{sec:bg_model}).
Background: exponential in $Q^2$ with $30{,}000$~fb total
cross section after veto cuts ($S/B \approx 3 \times 10^{-4}$).}
\label{fig:helicity}
\end{figure}

Figure~\ref{fig:helicity} shows the result.
With $500\fb^{-1}$ total luminosity split equally between helicity
states ($250\fb^{-1}$ per helicity), $\sim\!2{,}200$ CC events are
produced (all from LH electrons) alongside $\sim\!7.5 \times 10^6$
background events per helicity (signal-to-background ratio
$S/B \approx 3 \times 10^{-4}$).
The helicity subtraction recovers the CC signal within statistical
uncertainties. The $1/\sqrt{N_\LH + N_\RH}$ errors are driven by
the total event count in each bin, including both signal and
background.

We emphasize that the background model only affects Fig.~\ref{fig:helicity};
the Fisher projections in Sec.~\ref{sec:phase2} use signal-only
event counts and assume perfect background subtraction.

\section{Phase 2: Axial Form Factor Measurement}
\label{sec:phase2}

\subsection{$A_{UL}$ sensitivity to $\MA$}

\begin{figure*}[t]
\centering
\includegraphics[width=\textwidth]{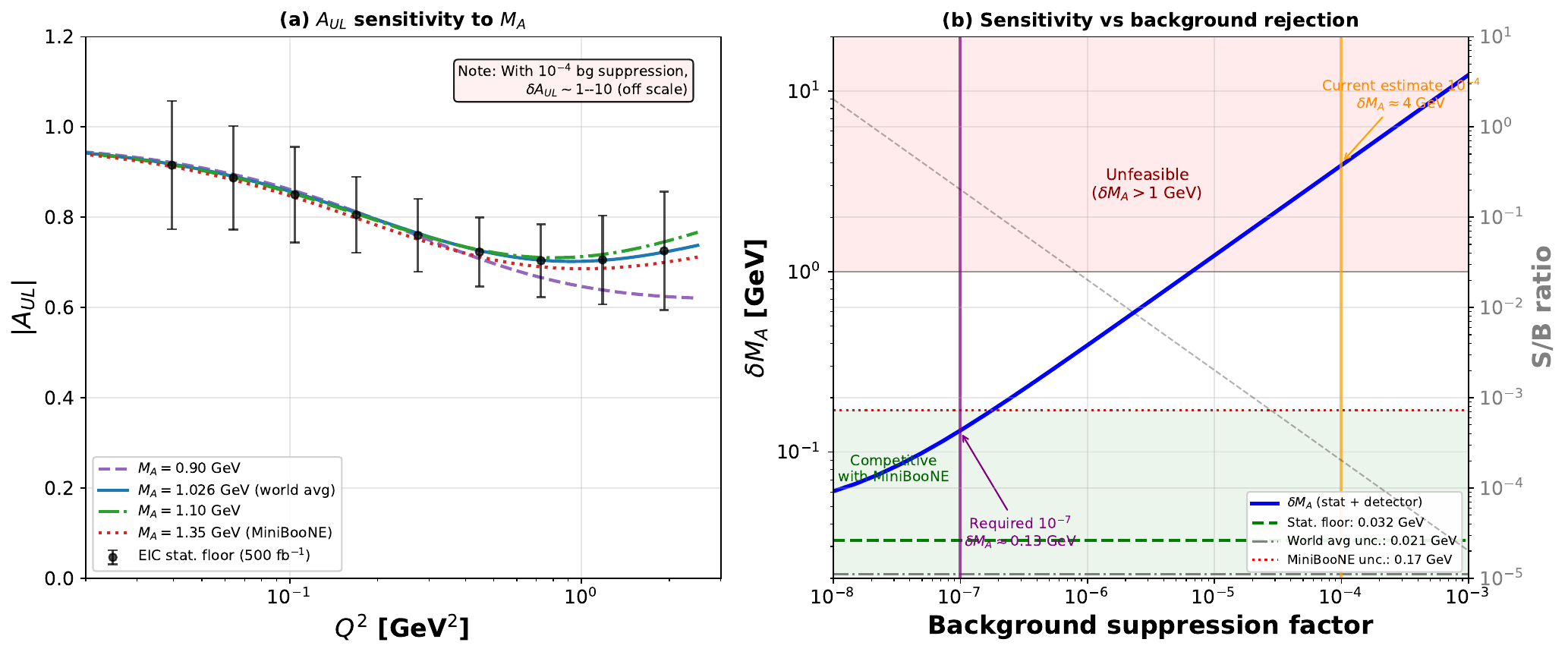}
\caption{(a)~Target-spin asymmetry $|A_{UL}|$ vs $Q^2$ for different
$\MA$ values. The curves separate in the sensitivity window
$0.1 < Q^2 < 1\GeV^2$. Black circles: ideal statistical floor.
With $10^{-4}$ background suppression, the asymmetry uncertainties
are $\delta A_{UL} \sim 1$--$10$ (off scale), providing no constraint.
(b)~Projected $\delta\MA$ vs background suppression factor over the
full range. The measurement is severely background-limited:
the current baseline ($10^{-4}$) yields $\delta\MA \approx 4\GeV$
(unfeasible). Achieving $\delta\MA \lesssim 0.17\GeV$ (MiniBooNE
precision) requires $10^{-7}$ suppression. The green and red shaded
regions indicate competitive and unfeasible regimes, respectively.}
\label{fig:AUL}
\end{figure*}

Figure~\ref{fig:AUL} shows $|A_{UL}(Q^2)|$ for four values of $\MA$:
0.90, 1.03, 1.10, and 1.35~GeV.
At low $Q^2$, all curves converge to $|A_{UL}| \approx 1$. In the
sensitivity window $Q^2 \sim 0.1$--$1\GeV^2$, the curves separate:
larger $\MA$ produces a flatter $F_A(Q^2)$, keeping $|A_{UL}|$ high
to larger $Q^2$.

The error bars show the projected EIC statistical precision from
Eq.~\eqref{eq:delta_AUL}, evaluated in 9 logarithmically spaced
$Q^2$ bins from 0.03 to 2.5~GeV$^2$.
Black circles show the ideal case ($\sim\!2{,}200$ events, no
background), with per-bin precision $\delta A_{UL} \sim 0.06$--$0.15$.
With the corrected $S/B \approx 3 \times 10^{-4}$, the realistic
asymmetry uncertainty is $\delta A_{UL} \sim 1$--$10$---far exceeding
$|A_{UL}| \lesssim 1$ and providing essentially no constraint.
The asymmetry measurement would only become useful with background
suppression approaching $10^{-7}$.

\subsection{Fisher matrix projection}
\label{sec:fisher}

We project the precision on $\boldsymbol{\theta} = (\gA, \MA)$
using two information sources, as described in
Sec.~\ref{sec:fisher_formalism}.

\paragraph{Cross section shape
$(\mathcal{F}^{(\sigma)})$.}
The expected events $N_k(\gA, \MA)$ in 11 $Q^2$ bins are computed
from Eq.~\eqref{eq:event_counting}. The sensitivity enters through
the $Q^2$ dependence: increasing $\gA$ raises $F_A$ at all $Q^2$
(changing the normalization), while increasing $\MA$ flattens the
$Q^2$ slope (changing the shape). Numerical derivatives
$\partial N_k / \partial\gA$ and $\partial N_k / \partial\MA$ are
evaluated by re-computing the full Llewellyn Smith cross section at
shifted parameter values.

\paragraph{Target-spin asymmetry
$(\mathcal{F}^{(A)})$.}
$A_{UL}$ is computed from the full Dirac trace at each shifted
$(\gA, \MA)$ point. The asymmetry is a \emph{ratio} of
spin-dependent cross sections, so it is insensitive to the overall
normalization but directly constrains the relative V--A interference.

\paragraph{Combined result.}
The total Fisher matrix
$\mathcal{F} = \mathcal{F}^{(\sigma)} + \mathcal{F}^{(A)}$ is
inverted to give the projected uncertainties
(Table~\ref{tab:fisher_results}).

\begin{table}[t]
\centering
\caption{Projected 1$\sigma$ uncertainties on $(\gA, \MA)$ from the
elastic CC channel with $500\fb^{-1}$ and $P_p = 0.70$. The ideal
projection uses signal-only Poisson statistics (Cram\'{e}r--Rao
floor). The realistic projection includes ZDC acceptance, $Q^2$
smearing, background noise from helicity subtraction, and
systematic uncertainties.}
\label{tab:fisher_results}
\begin{tabular}{lccc}
\toprule
& $\delta\gA$ & $\delta\MA$ [GeV] & $\rho$ \\
\midrule
\multicolumn{4}{l}{\emph{Ideal (Cram\'{e}r--Rao floor, ${\sim}2{,}200$ events)}} \\
\quad$d\sigma/dQ^2$ only & 0.046 (3.6\%) & 0.033 (3.2\%) & $-0.80$ \\
\quad$A_{UL}$ only & 0.56 (44\%) & 0.27 (26\%) & $-0.77$ \\
\quad Combined & 0.045 (3.6\%) & 0.032 (3.2\%) & $-0.79$ \\
\midrule
\multicolumn{4}{l}{\emph{Realistic ($\sim\!1{,}400$ signal, $S/B\!\approx\!3 \times 10^{-4}$)}} \\
\quad Stat.+detector & 5.9 (470\%) & 3.9 (380\%) & $-0.84$ \\
\quad Systematic & 0.019 & 0.019 & --- \\
\quad \textbf{Total} & \textbf{5.9 (470\%)} & \textbf{3.9 (380\%)}
& $\mathbf{-0.84}$ \\
\multicolumn{4}{l}{\footnotesize (Measurement unfeasible with $10^{-4}$ background suppression)} \\
\bottomrule
\end{tabular}
\end{table}

The ideal result $\delta\MA = 0.032\GeV$ (3.2\%) is dominated by the
$Q^2$-binned cross section rather than the asymmetry. With only
$\sim\!2{,}200$ events, the per-bin asymmetry measurement has limited
statistical power, while the cross section shape provides a direct
measurement of $F_A^2$ in each bin.
\rev{Quantitatively, decomposing the Fisher information: the
cross-section shape contributes $\sim\!98\%$ of the total information
on $\MA$, while $A_{UL}$ contributes only $\sim\!2\%$. The asymmetry
measurement is therefore a complementary probe rather than the
primary sensitivity driver. The value of $A_{UL}$ lies in its
different systematic dependencies (it is a ratio, so luminosity
uncertainties cancel), providing a cross-check rather than the
dominant constraint.}

The \emph{realistic} projection degrades the sensitivity catastrophically,
from $\delta\MA = 0.032\GeV$ (statistical floor) to $\delta\MA \gg 1\GeV$.
The dominant effect is the background noise from helicity subtraction:
with $S/B \approx 3 \times 10^{-4}$, the effective variance per bin is
$S + 2B \approx 6{,}700\,S$, degrading the Fisher information by
nearly four orders of magnitude. The ZDC acceptance loss (reducing events from
$\sim\!2{,}200$ to $\sim\!1{,}440$) and $Q^2$ smearing have modest
additional impact.
Systematic uncertainties total $\delta\MA^\text{syst} = 0.019\GeV$,
dominated by the ZDC energy scale (0.012~GeV) and residual background
subtraction (0.015~GeV), and are subdominant to the statistical+detector
contribution.

Improving the background suppression by orders of magnitude is essential:
\begin{itemize}
\item $10^{-5}$: $\delta\MA \approx 1.3\GeV$ (still unfeasible)
\item $10^{-6}$: $\delta\MA \approx 0.4\GeV$ (marginal)
\item $10^{-7}$: $\delta\MA \approx 0.14\GeV$ (competitive)
\end{itemize}
Achieving $10^{-7}$ suppression (three orders of magnitude beyond baseline)
represents an extraordinary experimental challenge.

\rev{Regarding the physics reach: the MiniBooNE anomaly corresponds
to a shift of $\Delta\MA = 1.35 - 1.03 = 0.32\GeV$. With the
statistical floor $\delta\MA = 0.032\GeV$, an EIC measurement
(if backgrounds can be sufficiently suppressed)
centered on the world average would disfavor the MiniBooNE value
at $0.32/0.032 \approx 10\sigma$. With the more realistic $10^{-7}$
suppression ($\delta\MA = 0.14\GeV$), this becomes
$0.32/0.14 \approx 2.3\sigma$. The key value of the EIC measurement is not merely
the precision, but that it is performed on a \emph{free proton},
eliminating the nuclear model uncertainties that complicate the
interpretation of the MiniBooNE result.}

\begin{figure}[t]
\centering
\includegraphics[width=\columnwidth]{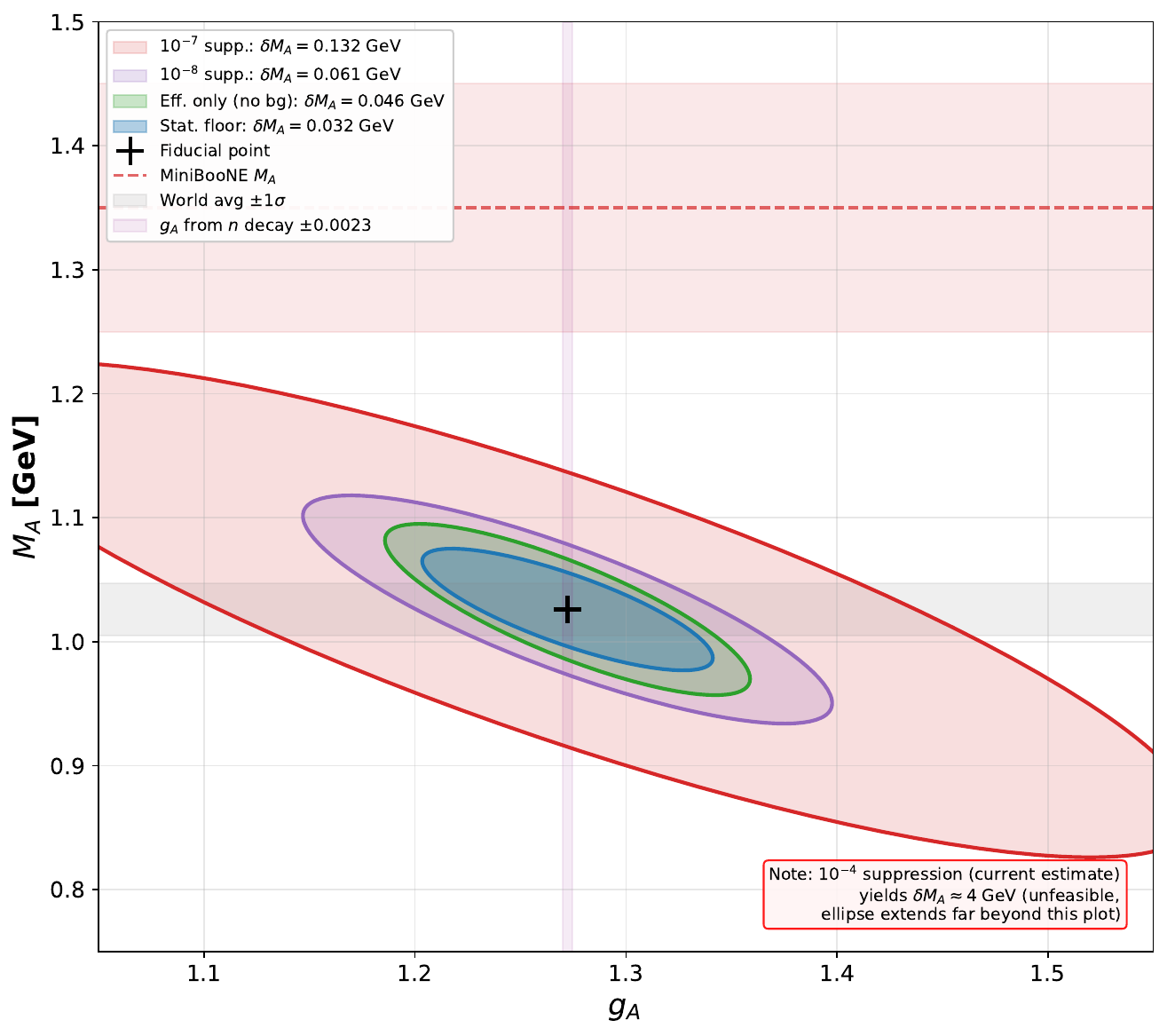}
\caption{Projected $1\sigma$ confidence contours in the $(\gA, \MA)$
plane for achievable suppression scenarios.
Blue: statistical floor (Cram\'{e}r--Rao bound, $\delta\MA = 0.032\GeV$).
Green: efficiency only, no background ($\delta\MA = 0.045\GeV$).
Purple: aggressive R\&D scenario with $10^{-8}$ suppression.
Red: minimum viable $10^{-7}$ suppression ($\delta\MA \approx 0.13\GeV$).
The current baseline ($10^{-4}$) yields $\delta\MA \approx 4\GeV$---the
ellipse would extend far beyond the plot and is not shown.
The purple vertical band shows the external $\gA$ constraint from
neutron beta decay.
Contours use $\Delta\chi^2 = 2.30$ for two parameters.}
\label{fig:ellipse}
\end{figure}

Figure~\ref{fig:ellipse} shows the confidence ellipses for different
background suppression scenarios. The ellipses exhibit a negative
correlation between $\gA$ and $\MA$: the cross section scales as
$\sigma \propto \gA^2$ at low $Q^2$, while the $Q^2$ shape depends
on $\MA$ through $(1 + Q^2/\MA^2)^{-4}$.
A simultaneous increase in $\gA$ and decrease in $\MA$ (or vice versa)
can produce similar cross sections over the measured $Q^2$ range,
hence the tilted ellipse.

The coupling $\gA = F_A(0) = 1.2723 \pm 0.0023$ is independently
determined from neutron beta decay~\cite{PDG2024} with 0.2\%
precision---far better than any scattering experiment can achieve.
Previous CCQE measurements therefore fix $\gA$ to this external value
and fit only $\MA$ from the $Q^2$ shape. The projected $\delta\MA$
values in Table~\ref{tab:fisher_results} follow this approach: they
correspond to projecting the 2D ellipse onto the $\MA$ axis using the
external $\gA$ constraint. In Figure~\ref{fig:ellipse}, the purple
band shows the $\gA$ constraint from neutron decay; the intersection
with each ellipse determines the effective $\MA$ uncertainty.

\rev{We emphasize that the EIC is not intended to measure $\gA$ more
precisely than neutron beta decay. Rather, the analysis uses $\gA$
as an external input (a Bayesian prior) to extract $\MA$---the same
approach used by all neutrino experiments. The value of the EIC
measurement is that it provides a consistency check: if the
beta-decay $\gA$, when used in CC scattering on a free proton, yields
an $\MA$ consistent with the deuterium world average, this confirms
the Standard Model V--A structure across different processes.
Conversely, a significant discrepancy would signal new physics or
unaccounted systematic effects.
We acknowledge that this consistency check is not fully independent:
since $\gA$ is used as an input to extract $\MA$, the two parameters
cannot be determined simultaneously with competitive precision.
The measurement tests whether the $(\gA, \MA)$ values from beta decay
and CC scattering are mutually consistent, rather than providing
independent determinations of both.}

Regarding the background rejection requirement: the baseline
$10^{-4}$ suppression yields $S/B \approx 3 \times 10^{-4}$,
making the measurement unfeasible. To achieve $S/B \approx 0.1$
and $\delta\MA \approx 0.14\GeV$, a background suppression of
$\sim\!10^{-7}$ would be required---three orders of magnitude
beyond current projections. We identify aggressive background
reduction as the critical challenge for this measurement.

\section{Phase 3: Structure Function Extraction}
\label{sec:phase3}

\subsection{The $y$-distribution method}

The principle underlying $xF_3$ extraction is similar to
Rosenbluth separation in elastic electron scattering: the cross section
is a linear combination of two unknown quantities ($G_E$ and $G_M$, or
here $F_2$ and $xF_3$), each multiplied by a known kinematic function.
By measuring at multiple kinematic points where the coefficients differ,
one can disentangle the two contributions.

In CC DIS, the role of the scattering angle is played by the
inelasticity $y$. At fixed Bjorken $x$, the cross section~\eqref{eq:cc_dis}
is \emph{linear} in $F_2$ and $xF_3$:
\begin{equation}
N_i = \alpha_i\left[Y_+(y_i)\,F_2 - Y_-(y_i)\,xF_3\right],
\label{eq:linear_model}
\end{equation}
where $\alpha_i = \frac{1}{2}\mathcal{L}\,P(y_i)\,\Delta x\,\Delta y_i$
contains the known prefactors including the $W$ propagator
$P(y) = (\GF^2 s/4\pi)(M_W^2/(Q^2+M_W^2))^2$.
The functions $Y_+ = 1 + (1-y)^2$ and $Y_- = 2y - y^2$ have
different $y$-shapes: $Y_+$ peaks at both $y = 0$ and $y = 1$,
while $Y_-$ vanishes at $y = 0$ and peaks at $y = 1$.
Fitting the observed $y$-distribution across many bins separates
the $F_2$ and $xF_3$ contributions, analogous to how varying the
electron scattering angle separates $G_E$ and $G_M$ in elastic scattering.

The Fisher matrix for $\boldsymbol{\theta} = (F_2, xF_3)$ at fixed
$x$ has analytical derivatives:
\begin{equation}
\frac{\partial N_i}{\partial F_2} = \alpha_i\,Y_{+,i}\,, \quad
\frac{\partial N_i}{\partial(xF_3)} = -\alpha_i\,Y_{-,i}\,,
\end{equation}
giving the $2\times 2$ Fisher matrix:
\begin{equation}
\mathcal{F} = \sum_i \frac{1}{N_i}
\begin{pmatrix}
(\alpha_i Y_{+,i})^2 & -\alpha_i^2 Y_{+,i} Y_{-,i} \\
-\alpha_i^2 Y_{+,i} Y_{-,i} & (\alpha_i Y_{-,i})^2
\end{pmatrix}.
\label{eq:fisher_dis}
\end{equation}

The structure functions $F_2$ and $xF_3$ are computed from the toy
PDF parameterization (Appendix~\ref{app:pdfs}). The absolute values
are approximate; the relevant feature for the sensitivity estimate is
the $xF_3/F_2$ \emph{ratio}, which determines how strongly the
$Y_-$ term perturbs the $y$-distribution.

The CC DIS cross section is large: $\sigma_\CC(Q^2 > 1\GeV^2)
\approx 105\pb$, yielding $\sim\!2.6 \times 10^7$ events with
$500\fb^{-1}$. Even in fine $(x, y)$ bins, the per-bin statistics are
$\order{10^4}$--$\order{10^5}$.

\subsection{Results}

\begin{figure*}[t]
\centering
\includegraphics[width=\textwidth]{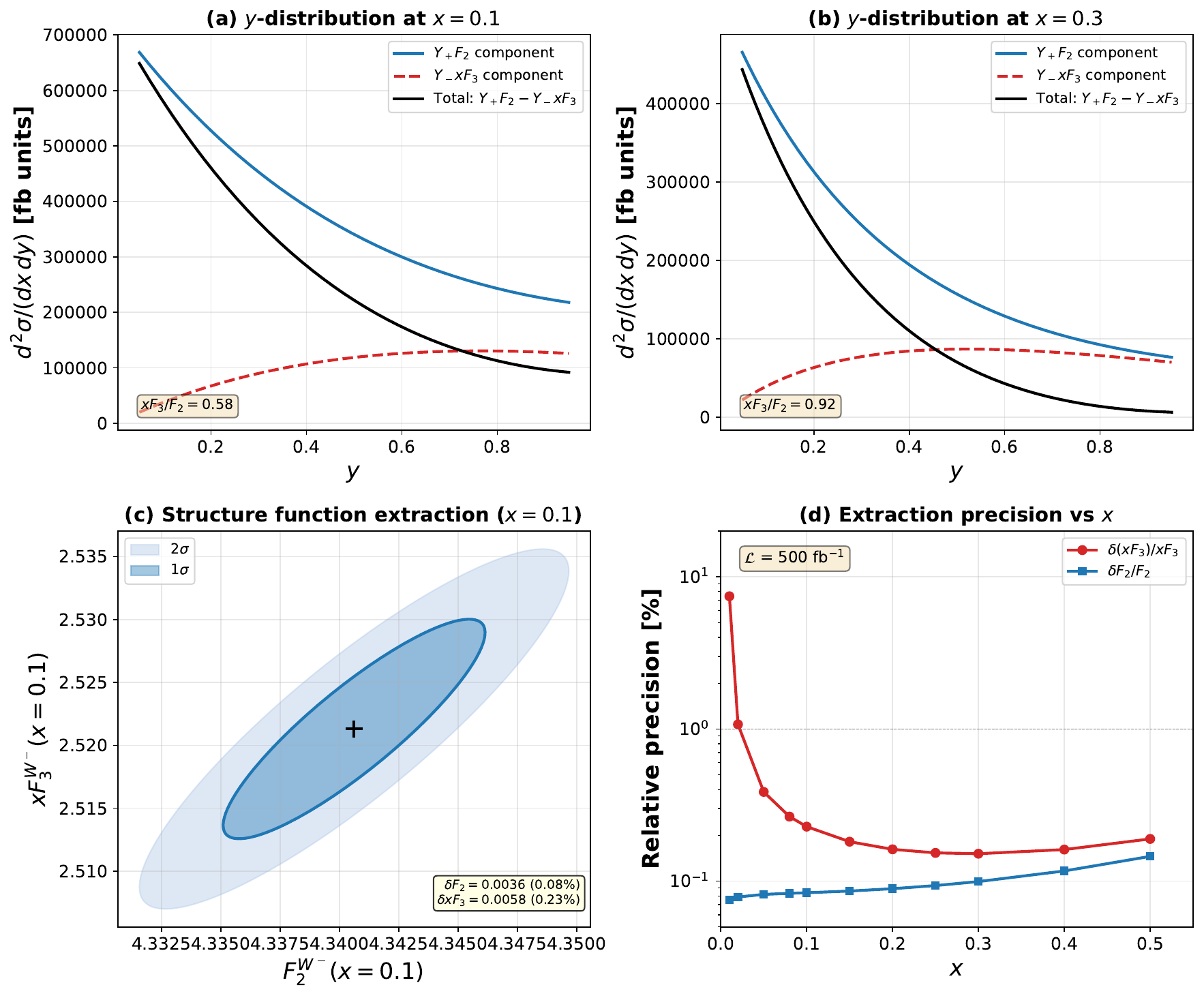}
\caption{CC DIS $y$-leverage for $xF_3$ extraction (all projections:
$500\fb^{-1}$, statistical only).
\emph{Upper:} $y$-distributions at $x = 0.1$ (left) and $x = 0.3$
(right), decomposed into $Y_+ F_2$ (blue) and $Y_- xF_3$ (red
dashed) components. Structure functions from toy PDFs
(Appendix~\ref{app:pdfs}).
\emph{Lower left:} Fisher-projected confidence ellipse for
$(F_2, xF_3)$ at $x = 0.1$ in 19 $y$-bins.
\emph{Lower right:} Relative extraction precision vs $x$, showing
sub-percent $xF_3$ extraction for $x > 0.05$.}
\label{fig:dis}
\end{figure*}

Figure~\ref{fig:dis} shows the $y$-distributions at $x = 0.1$ and
$x = 0.3$, decomposed into $Y_+ F_2$ and $Y_- xF_3$ components.
At $x = 0.1$, $xF_3/F_2 = 0.58$ (significant sea);
at $x = 0.3$, $xF_3/F_2 = 0.92$ (valence-dominated).

The lower-right panel shows the projected extraction precision vs
$x$.
$F_2$ is constrained to $\sim\!0.1\%$ over most of the $x$ range.
$xF_3$ is extracted to 0.2--0.4\% for $x > 0.05$, but degrades to
$\sim\!7\%$ at $x = 0.01$ where $xF_3/F_2 \to 0$.
At small $x$ the sea dominates, $xF_3 \approx 0$, and the $xF_3$
signal becomes a tiny perturbation on the dominant $Y_+ F_2$ term.

\rev{\paragraph{Theoretical limitations.}
The sub-percent statistical precision quoted above represents the
Cram\'{e}r--Rao floor, not the achievable experimental precision.
In practice, the extraction is limited by:
(i)~QED radiative corrections, which distort the $y$-distribution
at the few-percent level and must be unfolded using Monte Carlo;
(ii)~NLO QCD corrections to the structure functions and the
Callan--Gross relation ($F_L \neq 0$);
(iii)~higher-twist effects at moderate $Q^2$.
A realistic estimate, accounting for these theoretical systematics,
is $\delta(xF_3)/xF_3 \sim 2$--$5\%$ (Table~\ref{tab:comparison}).
The statistical floor demonstrates that the EIC has ample event
rates; the challenge is controlling the theoretical uncertainties.}

\rev{\paragraph{NC DIS background.}
The main experimental background for CC DIS is neutral-current (NC)
DIS where the scattered electron is lost (undetected in a crack,
down the beam pipe, or misidentified as a hadron).
For $Q^2 > 1\GeV^2$, the electron scattering angle is
$\theta_e \gtrsim 4^\circ$, well within the ePIC central detector
acceptance, so electron detection efficiency exceeds 99\%.
With $\sigma_\NC / \sigma_\CC \sim 3$--$5$, the raw NC
contamination is a few percent.
This can be further suppressed by requiring large missing transverse
momentum: CC events have $p_T^\text{miss} \sim 10$--$50\GeV$
(the neutrino), while NC events with a lost electron have
$p_T^\text{miss} \sim \sqrt{Q^2} \sim 1$--$5\GeV$.
After a $p_T^\text{miss} > 10\GeV$ cut, the residual NC
contamination is $\lesssim 1\%$, yielding $S/B \gtrsim 100$.
This contrasts sharply with the elastic channel, where
$S/B \sim 10^{-4}$ before any kinematic cuts.
The NC background dilutes the $xF_3$ signal (since NC has no
parity-violating $xF_3$ from photon exchange), but the
sub-percent contamination level is smaller than the theoretical
systematics and can be corrected using measured electron veto
efficiencies.}

\section{Discussion}
\label{sec:discussion}

\subsection{Comparison with neutrino experiments}

Table~\ref{tab:comparison} compares the projected EIC measurements
with existing experiments.

\begin{table}[t]
\centering
\caption{Comparison of $\MA$ and $xF_3$ measurements. EIC
projections include both the ideal statistical floor and the
realistic estimate with detector effects. NuTeV and CHORUS
uncertainties include full systematic and nuclear corrections.}
\label{tab:comparison}
\begin{tabular}{llll}
\toprule
Experiment & Target & $\delta\MA$ & Note \\
\midrule
D$_2$ bubble ch.\ & Free nucleon & $\pm 0.021$ GeV & World avg \\
MiniBooNE & Carbon & $\pm 0.17$ GeV & $\MA\!\approx\!1.35$ GeV \\
MINERvA & Hydrogen & -- & Limited stats \\
\textbf{EIC (ideal)} & \textbf{Free $p$} & $\pm\mathbf{0.032}$ \textbf{GeV}
& \textbf{Stat.\ floor} \\
\textbf{EIC (stat.\ floor)} & \textbf{Free $p$} & $\pm\mathbf{0.032}$ \textbf{GeV}
& \textbf{Ideal} \\
\midrule
 & & $\delta(xF_3)/xF_3$ & \\
\midrule
NuTeV~\cite{NuTeV2006} & Iron & 5--10\% & Full syst.\ \\
CHORUS~\cite{CHORUS2005} & Lead & $\sim\!10\%$ & Full syst.\ \\
\textbf{EIC (ideal)} & \textbf{Free $p$} & $\mathbf{<1\%}$
& \textbf{Stat.\ only} \\
\textbf{EIC (realistic)} & \textbf{Free $p$} & $\mathbf{2\text{--}5\%}$
& \textbf{Est.\ syst.}$^\dag$ \\
\bottomrule
\end{tabular}
\vspace{1ex}

{\footnotesize $^\dag$Estimated including radiative corrections and
detector resolution; requires full Monte Carlo study.}
\end{table}

The EIC measurements have two fundamental advantages:
(1)~a \emph{free proton target}, eliminating the nuclear model
uncertainties that affect all current precise $xF_3$ measurements; and
(2)~a \emph{known beam}, eliminating the 5--10\% neutrino flux
uncertainties.

\subsection{Systematic uncertainties and their impact}
\label{sec:systematics}

Table~\ref{tab:syst_budget} summarizes the systematic uncertainty
budget for the realistic $\MA$ projection. The total systematic
$\delta\MA^\text{syst} = 0.019\GeV$ would be subdominant to the
statistical floor of $0.032\GeV$ in the ideal case. However, with realistic
background levels ($S/B \approx 10^{-4}$), systematic uncertainties are
entirely negligible. The methodology for
propagating each source is described in Sec.~\ref{sec:detector_model}.

\begin{table}[t]
\centering
\caption{Systematic uncertainty budget for $\delta\MA$ in the
realistic projection.}
\label{tab:syst_budget}
\begin{tabular}{lcc}
\toprule
Source & Assumption & $\delta\MA$ [GeV] \\
\midrule
Background subtraction & 1\% LH/RH asymmetry & 0.015 \\
ZDC energy scale & 1\% & 0.012 \\
Luminosity & 2\% & 0.004 \\
Polarimetry ($P_p$) & 1\% & 0.001 \\
\midrule
\textbf{Total (quadrature)} & & \textbf{0.019} \\
\bottomrule
\end{tabular}
\end{table}

\paragraph{Hierarchy of uncertainties.}
The dominant uncertainty source is statistical noise from the helicity
subtraction ($S + 2B$ variance), followed by ZDC acceptance loss.
Systematic uncertainties are currently subdominant. Even with $10^{-7}$
background suppression (yielding $\delta\MA \approx 0.14\GeV$),
the ZDC energy scale and background subtraction would
each contribute $\sim\!0.01$--$0.02\GeV$, remaining subdominant.

\paragraph{Paths to improvement.}
The background subtraction systematic could be reduced by improved
LH/RH luminosity monitoring or by using the RH data to directly
constrain the background $Q^2$ shape (currently we assume the shape
is known). The ZDC energy scale could be calibrated using exclusive
$\pi^0 \to \gamma\gamma$ decays or the $\eta \to \gamma\gamma$ peak.
Radiative corrections can be computed with existing tools (RADCOR)
and unfolded.

\paragraph{Form factor model dependence.}
The dipole \emph{ansatz} for $F_A(Q^2)$ is conventional but not
fundamental. Model-independent extractions using the
z-expansion~\cite{BernardMA} parameterization could be performed
with the same data, at the cost of additional fit parameters.
The vector form factor uncertainties (1--2\%~\cite{Arrington2007})
contribute a small additional systematic not included in
Table~\ref{tab:syst_budget}.

\subsection{Nuclear targets}
\label{sec:nuclear}

The EIC will collide electrons with nuclear beams (d, He, C, Ca, Pb),
enabling a direct measurement of the $A$-dependence of CC scattering.
This could explain the MiniBooNE anomaly as a nuclear effect.

To illustrate one component of nuclear effects, we apply a local Fermi
gas model~\cite{SmithMoniz1972} with Pauli blocking.
In a nucleus, the final-state neutron must have momentum above the
Fermi surface ($k_F \approx 225\MeV$ for $A \geq 12$).
This blocks low-$Q^2$ events where the three-momentum transfer
$|\vec{q}\,| < 2k_F$, preferentially removing events where $F_A$ is
largest.

\begin{figure*}[t]
\centering
\includegraphics[width=\textwidth]{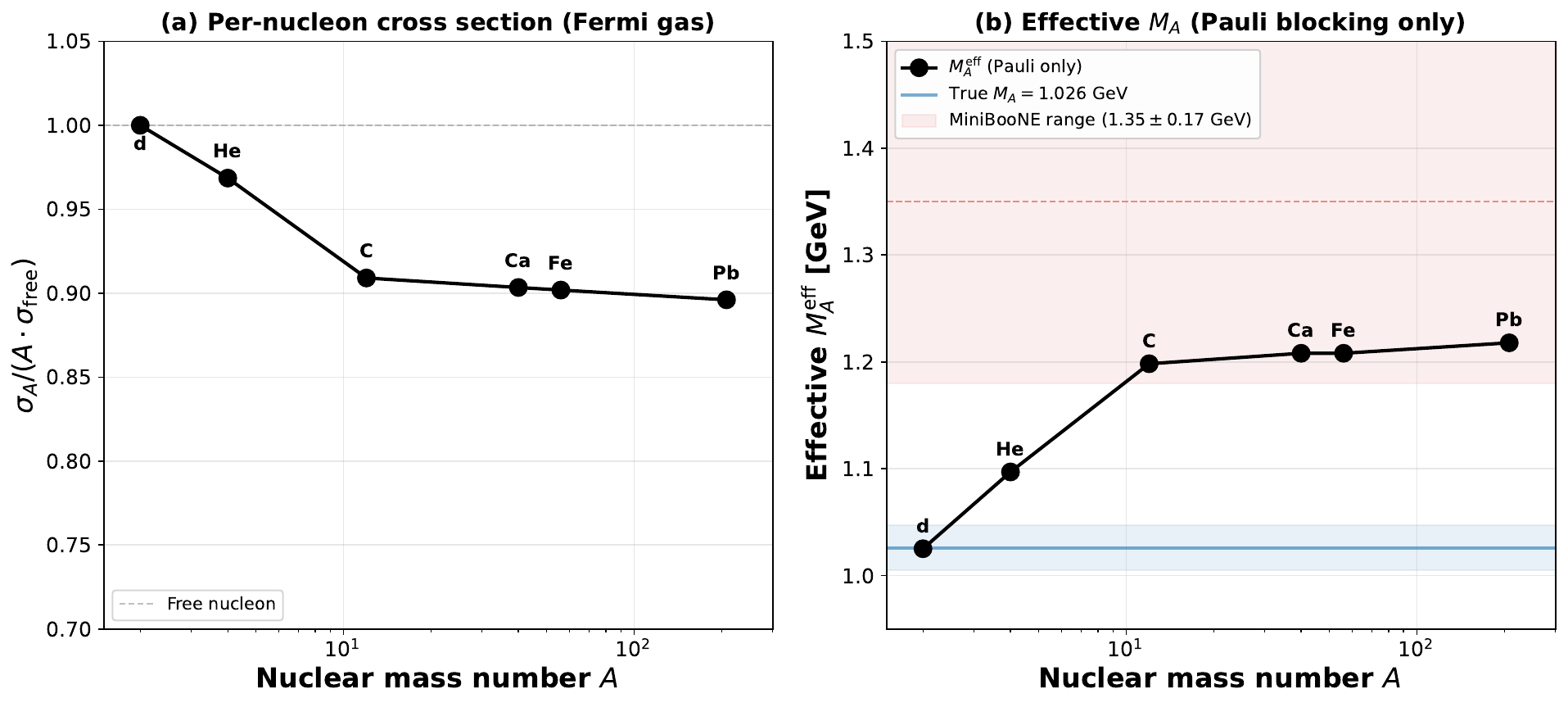}
\caption{Illustrative nuclear target scaling using a simplified
local Fermi gas model~\cite{SmithMoniz1972} with Pauli blocking only.
\emph{Left:} Per-nucleon cross section ratio
$\sigma_A/(A \cdot \sigma_\text{free})$ vs $A$, showing Pauli
blocking suppression.
\emph{Right:} Effective $\MA^\text{eff}$ extracted by fitting the
nuclear-modified $Q^2$ shape with the free-nucleon formula. The blue
band is the true $\MA = 1.026\GeV$; the red band is the MiniBooNE
range.
Note: This simplified model shows only Pauli blocking; the full
MiniBooNE anomaly is primarily attributed to 2p2h/MEC effects
(Sec.~\ref{sec:nuclear}), which produce larger shifts than
shown here.}
\label{fig:nuclear}
\end{figure*}

Figure~\ref{fig:nuclear} shows the result.
The per-nucleon cross section decreases with $A$ due to Pauli blocking
(left panel).
When the nuclear-modified $Q^2$ distribution is fit with the free-nucleon
formula, the extracted $\MA^\text{eff}$ exceeds the true value
(right panel): Pauli blocking removes the low-$Q^2$ events that
anchor $F_A$, mimicking a harder (more slowly falling) form factor.
For carbon ($A = 12$), Pauli blocking alone shifts $\MA^\text{eff}$
upward by $\sim\!0.1$--$0.2\GeV$.

We emphasize that this is a simplified illustration. The modern
consensus~\cite{Martini2009, Nieves2011} attributes the MiniBooNE
anomaly primarily to \emph{two-particle--two-hole (2p2h)} excitations
and meson exchange currents, which are not included in the simple
Fermi gas model. These multinucleon correlations produce additional
cross section enhancement at moderate $Q^2$ that mimics CCQE events
and further inflates $\MA^\text{eff}$. A full treatment requires
sophisticated many-body calculations beyond the scope of this paper.
The key point is that measurements on nuclear targets cannot
straightforwardly extract the free-nucleon $\MA$. In contrast, the
EIC provides a clean probe on a free proton.

\subsection{Caveats and future improvements}
\label{sec:caveats}

This sensitivity study relies on several approximations that should
be addressed in future work:

\paragraph{Background model.}
The photoproduction background is modeled as an exponential
$dN/dQ^2 \propto \exp(-Q^2/\Lambda^2)$ with $\Lambda^2 = 0.3\GeV^2$,
based on pion-exchange phenomenology.
The $10^{-4}$ suppression factor is from Ref.~\cite{Klest2025};
the actual achievable suppression depends critically on the
forward veto design and hadron rejection algorithms.
The veto strategy relies on detecting the $\pi^+$ from
$\gamma^* p \to n\,\pi^+\,X$ in the forward tracker; at very low
$Q^2$ (quasi-real photoproduction), the scattered electron remains
in the beam pipe and cannot be used for rejection.
Additionally, \emph{beam-gas backgrounds}---electrons scattering off
residual gas in the vacuum pipe---can produce fake CC signatures
(neutrons and missing energy) and are not included in this analysis.
A full GEANT4 simulation with realistic detector response and vacuum
conditions is needed to validate these estimates.

\paragraph{ZDC acceptance.}
We use a simplified ZDC model with a plateau efficiency of 85\%
for $\theta_n < 3.5$~mrad and linear falloff to zero at 5.5~mrad.
The actual acceptance depends on the beam optics, magnet apertures,
and hadronic shower containment. The beam-pipe hole at $\theta = 0$
may affect the lowest-$Q^2$ bin.

\paragraph{Dipole form factor assumption.}
The analysis assumes $F_A(Q^2) = \gA/(1 + Q^2/\MA^2)^2$. The z-expansion
formalism~\cite{BernardMA} allows model-independent extraction with
more parameters; this would increase the uncertainty but provide
a more robust result. The dipole assumption may introduce bias if
the true form factor deviates from this form.

\paragraph{CC DIS analysis.}
The CC DIS $y$-leverage analysis (Phase~3) uses ideal Fisher
information with no detector smearing or background. A full
treatment would require $(x, y, Q^2)$ response matrices and
misidentified NC backgrounds.

\paragraph{Radiative corrections.}
QED radiative corrections modify the effective $Q^2$ and $y$
distributions at the few-percent level. For the elastic channel,
the corrections are $\order{\alpha/\pi} \sim 0.2\%$ and negligible.
For the CC DIS analysis, radiative effects can be more substantial
(several percent) due to initial-state radiation shifting the
effective beam energy. The sub-percent $xF_3$ precision claimed in
Table~\ref{tab:comparison} is \emph{statistical only}; a full
treatment would require radiative corrections unfolded with
high-fidelity Monte Carlo, which could degrade the precision to the
few-percent level. This is comparable to the NuTeV/CHORUS systematics
and represents a significant caveat for the DIS analysis.

\paragraph{Path forward.}
The key experimental challenge is background rejection.
Our analysis identifies $10^{-7}$ suppression as the target for a
competitive $\MA$ measurement ($\delta\MA \approx 0.14\GeV$).
This is three orders of magnitude beyond current projections.
Achieving this requires:
(1)~multi-layer forward charged-particle veto with $>99.9\%$
rejection per layer,
(2)~ZDC timing and position cuts to reject non-elastic
topologies, and
(3)~machine-learning algorithms trained on PYTHIA/GEANT4 samples
to discriminate CC elastic from photoproduction.
With these improvements, the EIC could achieve world-leading
precision on the nucleon axial form factor using a clean,
nuclear-effect-free probe.

\section{Conclusions}
\label{sec:conclusions}

We have demonstrated that the Electron-Ion Collider has the
statistical reach to make significant contributions to neutrino
physics through charged-current electron--proton scattering.

Using the Fisher information formalism with the inputs in
Tables~\ref{tab:eic_params}--\ref{tab:inputs}, we present both
the Cram\'{e}r--Rao statistical floor (ideal) and a first-order
realistic projection incorporating ZDC acceptance, $Q^2$ smearing,
background noise, and systematic uncertainties
(Table~\ref{tab:fisher_results}):

\begin{enumerate}
\item \textbf{Helicity filtering} provides clean CC signal
extraction using RH electron data as an \emph{in situ} EM background
template, yielding $S/B \approx 3 \times 10^{-4}$ after forward veto cuts
and $\sim\!2{,}200$ CC events from $500\fb^{-1}$
(Fig.~\ref{fig:helicity}).

\item \textbf{The elastic CC cross section shape} yields a
Cram\'{e}r--Rao floor of $\delta\MA = 0.032\GeV$ (3.2\%).
With realistic detector effects, this degrades to
$\delta\MA \gg 1\GeV$, dominated by background noise
from the helicity subtraction ($S/B \approx 3 \times 10^{-4}$, reducing the
effective Fisher information by $\sim\!6{,}700\times$).
Achieving competitive sensitivity ($\delta\MA \approx 0.14\GeV$)
would require background suppression of $\sim\!10^{-7}$---three
orders of magnitude beyond current projections
(Fig.~\ref{fig:ellipse}, Table~\ref{tab:fisher_results}).

\item \textbf{The CC DIS $y$-distribution} separates
$F_2^{W^-}$ and $xF_3^{W^-}$ with sub-percent statistical precision
for $x > 0.05$, providing the first measurement of $xF_3$ on a free
proton (Fig.~\ref{fig:dis}). This measurement does not suffer from
the leading-neutron photoproduction background that limits the
elastic channel, making it the most promising near-term
electroweak physics goal at the EIC.
\end{enumerate}

The key advantage of the EIC is the \emph{free proton target},
which eliminates the nuclear model uncertainties that dominate
current $\MA$ and $xF_3$ extractions.
If sufficient background suppression ($\sim\!10^{-7}$) can be achieved,
a measurement of $\MA$ at the 14\% level on hydrogen would
provide qualitatively different information from the $\pm 0.17\GeV$
MiniBooNE measurement on carbon, as it directly constrains the
nucleon-level form factor without nuclear corrections.

The severely background-limited nature of the elastic channel measurement
identifies a critical challenge: current projections ($10^{-4}$ suppression)
yield an unfeasible $\delta\MA \approx 4\GeV$. Dedicated R\&D on
background rejection strategies (improved forward tracking, neutral particle
identification, machine-learning event classification) is essential.
The elastic channel remains an important long-term goal: the statistical
floor of $\delta\MA = 0.032\GeV$ demonstrates that the fundamental
physics reach exists if backgrounds can be controlled.
The nuclear target program (Sec.~\ref{sec:nuclear}) represents a
further extension that could directly map the $A$-dependence of
neutrino cross sections and definitively resolve the $\MA$ anomaly.

\begin{acknowledgments}
[Acknowledgments to be added.]
\end{acknowledgments}

\appendix

\section{Dirac trace technique}
\label{app:trace}

The squared matrix element $|\mathcal{M}|^2$, summed over final spins
and averaged over initial spins, is computed using:
\begin{equation}
\sum_\text{spins} |\bar{u}(p')\,\Gamma\,u(p)|^2
= \text{Tr}\!\left[(\slashed{p}' + m')\,\Gamma\,
(\slashed{p} + m)\,\bar{\Gamma}\right],
\end{equation}
where $\bar{\Gamma} = \gamma^0 \Gamma^\dagger \gamma^0$.
For the CC elastic process:
\begin{equation}
|\mathcal{M}|^2 = \frac{\GF^2 |\Vud|^2}{2}\, L_{\mu\nu}\, H^{\mu\nu}\,,
\end{equation}
with the leptonic tensor:
\begin{equation}
L^{\mu\nu} = \text{Tr}\!\left[\slashed{k}'\,\gamma^\mu(1-\gamma_5)\,
\slashed{k}\,(1+\gamma_5)\,\gamma^\nu\right],
\end{equation}
and the hadronic tensor (target spin four-vector $s$):
\begin{equation}
H^{\mu\nu} = \text{Tr}\!\left[(\slashed{P}'+M_n)\,\Gamma^\mu\,
(\slashed{P}+M_p)\,\frac{1+\gamma_5\slashed{s}}{2}\,
\bar{\Gamma}^\nu\right].
\end{equation}
In our implementation, each $\gamma^\mu$ is a $4\times 4$ complex
matrix in the Dirac representation, and the trace is computed
numerically.

\rev{\paragraph{Spin four-vector.}
For longitudinal polarization of a proton with four-momentum
$P = (E, 0, 0, |\vec{p}|)$ moving along the $z$-axis, the spin
four-vector is:
\begin{equation}
s^\mu = \frac{1}{M_p}\left(|\vec{p}|, 0, 0, E\right),
\end{equation}
which satisfies $s \cdot P = 0$ and $s^2 = -1$.
For the proton at rest ($|\vec{p}| = 0$), this reduces to
$s^\mu = (0, 0, 0, 1)$ (spin along $+z$).
The orthogonality condition $s \cdot P = 0$ is verified numerically
at each $Q^2$ point.}

\rev{\paragraph{Mass treatment.}
In the hadronic tensor, we use the physical proton mass $M_p$ for
the initial state and neutron mass $M_n$ for the final state.
For kinematic quantities (e.g., $\tau = Q^2/4M^2$), we use the
average nucleon mass $M = (M_p + M_n)/2 = 0.939\GeV$. The mass
difference $(M_n - M_p)/M \approx 0.14\%$ has negligible impact
on the extracted $\MA$.}

\rev{\paragraph{Pseudoscalar form factor.}
The hadronic vertex $\Gamma^\mu$ includes all four form factors:
$F_1$, $F_2$, $F_A$, and $F_P$. Although $F_P$ is suppressed by
$m_\ell^2/M^2 \sim 10^{-7}$ for electrons, we include it via the
PCAC relation (Eq.~\ref{eq:PCAC}) to ensure completeness. Its
contribution to both the unpolarized cross section and $A_{UL}$
is verified to be $< 10^{-6}$ relative.}

\rev{\paragraph{Validation.}}
This is verified against the Llewellyn Smith formula to
$d\sigma_\text{trace}/d\sigma_\text{LS} = 1.0000$ at all $Q^2$.
\rev{For the spin-dependent terms, we verify $A_{UL}(Q^2 \to 0) \to 1$
(the low-$Q^2$ limit where the axial contribution dominates) and
compare against the analytical asymmetry from
Ref.~\cite{Akbar2017}, finding agreement to $< 0.1\%$ across the
full $Q^2$ range.}

\section{Toy PDF parameterization}
\label{app:pdfs}

For the CC DIS analysis, we use analytical PDFs at a fixed scale
$Q^2 \sim 10\GeV^2$:
\begin{align}
x\,u_v(x) &= 3.0\, x^{0.5}\, (1-x)^{3.0}\, (1 + 3\sqrt{x})\,, \\
x\,d_v(x) &= 1.0\, x^{0.5}\, (1-x)^{4.0}\, (1 + 5\sqrt{x})\,, \\
x\,\bar{q}(x) &= 0.4\, x^{-0.2}\, (1-x)^{7.0}\,, \\
x\,c(x) &= 0.15\, (1-x)^{8.0}\,,
\end{align}
with $\bar{u} = \bar{d} = \bar{q}$ and
$s = \bar{s} = 0.5\,\bar{q}$.
These capture the qualitative features of modern PDF sets:
$u_v$ peaks at $x \approx 0.2$ with integral~2,
$d_v$ peaks at $x \approx 0.15$ with integral~1,
the sea rises at small $x$, and charm is suppressed.
DGLAP evolution is not included.
For a definitive analysis, LHAPDF~\cite{LHAPDF} with a modern PDF
set should be used; the toy parameterization is sufficient for
demonstrating the $y$-leverage method and estimating the statistical
precision.


\end{document}